\documentclass[amsmath,amssymb, aps,pra,twocolumn,superscriptaddress,nofootinbib,showkeys,a4paper]{revtex4-2}

\usepackage{orcidlink}
\usepackage{cancel}
\usepackage{datetime}
\usepackage{cancel}
\usepackage{braket}
\usepackage{siunitx}

\usepackage{booktabs}
\usepackage{graphicx}
\begin{document}\title{Broken-symmetry phenomena enhanced by quasi-bound states in the continuum}\author{Jan David Fischbach\orcidlink{0009-0003-8765-8920}
}\email{fischbach@kit.edu}
\affiliation{Karlsruhe Institute of Technology (KIT), Germany}\author{Lukas Rebholz\orcidlink{0009-0001-0737-7826}
}\affiliation{Karlsruhe Institute of Technology (KIT), Germany}\author{Nikita Ustimenko\orcidlink{0000-0002-5137-493X}
}\affiliation{Karlsruhe Institute of Technology (KIT), Germany}\author{Markus Nyman\orcidlink{0000-0002-9661-6461}
}\affiliation{Karlsruhe Institute of Technology (KIT), Germany}\author{Carsten Rockstuhl\orcidlink{0000-0002-5868-0526}
}\affiliation{Karlsruhe Institute of Technology (KIT), Germany}\author{Ivan Fernandez-Corbaton\orcidlink{0000-0003-2834-5572}
}\affiliation{Karlsruhe Institute of Technology (KIT), Germany}\newdate{articleDate}{16}{6}{2026}
\date{\displaydate{articleDate}}
\begin{abstract}
Many of the most powerful and elegant models in physics are grounded in symmetries. In electrodynamics, for example, geometric symmetries govern the observable effects of light-matter interactions. However, for man-made objects, exact symmetries are rarely met and tiny deviations are common. Nonetheless, even approximate symmetries keep many symmetry-derived rules effectively intact. However, as we will show here, this is not universally true.
We demonstrate that an incremental violation of the symmetry of a carefully designed system can produce an optical response maximally different from the unbroken symmetry case. To do so, we exploit symmetry-protected quasi-bound states in the continuum (qBICs). Specifically, we design a four-fold rotationally symmetric metasurface composed of nearly dual-symmetric meta-atoms that supports a pair of spectrally aligned electric and magnetic qBICs. At normal incidence, symmetry forbids helicity-preserving reflection. However, for arbitrarily small deviations from normal incidence, the strong resonant enhancement associated with the qBICs overcomes the near-symmetry suppression and enables perfect helicity-preserving reflection. This rapidly emerging violation of symmetry-rules reveals a fundamental intricacy when it comes to treating near-symmetric systems. At the same time, our work opens the door to novel applications in metrology and sensing.
\end{abstract}
\maketitle

\section{Introduction and Theoretical Background}

Symmetries are crucial to modern physics. Invariances under transformations directly dictate the fundamental forces~\citep{weinbergQuantumTheoryFields1995}, conserved quantities~\citep{noetherInvarianteVariationsprobleme1983}, and selection rules~\citep{weinbergQuantumTheoryFields1995, wignerGroupTheoryIts2012}, providing the very structure to our understanding of the laws of nature.
However, many interesting phenomena only arise once these invariances are violated. Or to put it in words similar to those of Wolfgang Pauli: God made the symmetries; their violation was invented by the devil.

While the binary question of whether a symmetry is broken or not is at the heart of many theories, approaches that quantify how strongly a given system breaks a symmetry have similarly attracted widespread interest. Examples include order parameters, measuring the magnitude of broken symmetries in liquid crystals~\citep{degennesPhysicsLiquidCrystals1993}, quantitative measures of duality and reciprocity breaking in photonics~\citep{fernandez-corbatonDualChiralObjects2015, kiselevTestingAccuracyConvergence2025}, and measures of CP violation in particle physics~\citep{jarlskogCommutatorQuarkMass1985, wolfensteinParametrizationKobayashiMaskawaMatrix1983}. Here, we present an example of a perturbative symmetry breaking, resulting in the immediate and maximal emergence of an otherwise forbidden effect, in the framework of classical linear electrodynamics. There, small perturbations to a symmetry typically generate only small violations of the rules which result from that symmetry~\citep{abdelrahmanEffectsSymmetrybreakingElectromagnetic2021}.

%  David Gross [@grossRoleSymmetryFundamental1996]: "The secret of nature is symmetry, but much of the texture of the world is due to mechanisms of symmetry breaking."

The phenomenon we discuss relies on the interplay of two symmetries: Electromagnetic duality and discrete rotational symmetry. The former describes an invariance under the interchange of the electric and magnetic fields. Sourceless Maxwell's equations in vacuum possess this \textbf{duality symmetry}~\citep{calkinInvariancePropertyFree1965}. Hence, we can separate them into decoupled equations for the $\pm1$ eigenvalues of the helicity operator $\mathbf{J} \cdot \mathbf{P}/|\mathbf{P}|$, which is the angular momentum $\mathbf{J}$ projected onto the direction of the linear momentum $\mathbf{P}$~\citep{tungGroupTheoryPhysics1985}. For plane waves, these helicity eigenvalues corresponds to left $\sigma_+$ and right circular $\sigma_ -$ polarizations.
However, the presence of electric charges breaks duality symmetry, coupling the otherwise separate helicities~\citep{zwanzigerQuantumFieldTheory1968}. Appropriate choices for the geometry of a scatterer allow restoring duality symmetry for the scattered field~\citep{fernandez-corbatonHelicityDualitySymmetry2015}. That is to say, when such a scatterer is illuminated with a field of pure helicity, the scattered field outside of scatterer has the same helicity. In this case, we speak of a dual scatterer, which requires a balanced electric and magnetic response.

At the same time, sufficient (at least three-fold) \textbf{discrete rotational symmetry} of the scatterer, under plane wave illumination along the rotation axis, dictates perfect helicity preservation in the forward scattered wave, while helicity flipping is enforced for backward scattering along the symmetry axis~\citep{fernandez-corbatonForwardBackwardHelicity2013}.

A scatterer that simultaneously possesses duality symmetry and sufficient rotational symmetry guarantees the complete suppression of backscattering, which can be understood as a generalized Kerker-condition~\citep{fernandez-corbatonForwardBackwardHelicity2013}.\footnote{The authors of Ref.~\citep{yangElectromagneticDualityProtected2020} have shown, that the symmetry requirements for the generalized Kerker-condition can be further relaxed by introducing a non-trivial permeability, which we will, however, not use here.} In the remainder of this article we present a metasurface that, despite possessing both duality and four-fold rotational symmetry, sustains extremely strong helicity-preserving backward scattering (\textit{i.e.}, reflection) as soon as the rotational symmetry is even slightly broken by off-axis illumination. The metasurface relies on quasi-bound states in the continuum (qBICs), which become increasingly resonant close to normal incidence, to mitigate the suppression of helicity-preserving reflection due to the four-fold rotational symmetry. In particular, we will use two qBICs centered at the same frequency, one with a dominant electric dipole and one with a dominant magnetic dipole contribution. This results in a polarization independent phase shift and will ensure the sought after helicity preservation. As we show in the appendix, for negligible non-radiative losses, temporal coupled mode theory predicts full reflection (given the z-mirror symmetry of our structure). In a separate section of the appendix, we investigate the effect of appreciable material loss, finding that even for realistic damping substantial helicity preserving reflection is sustained beyond angles smaller than a degree.

Helicity-preserving cavities offer a promising platform for enantiomer-selective probing and manipulation of biomolecules and other chiral matter. State-of-the-art designs rely on grazing incidence~\citep{nymanDigitalTwinChiral2024a, feisHelicityPreservingOpticalCavity2020} or low-symmetry lattices~\citep{pozharkovaLowsymmetryLatticesNonchiral2026} to realize the required helicity-preserving reflection. On top of being intellectually appealing, the effect we demonstrate here provides a potential alternative mechanism to realize mirrors for such cavities at near-normal incidence.

%  _The numerically observed effect provides a new mechanism to realize mirrors for helicity preserving cavities. Such cavities have the potential to enhance enantio-selective light-matter interactions, and currently rely on either ... ?_

\section{An (almost) dual-symmetric meta-atom}

\begin{figure}[!htbp]
\centering
\includegraphics[width=0.8\linewidth]{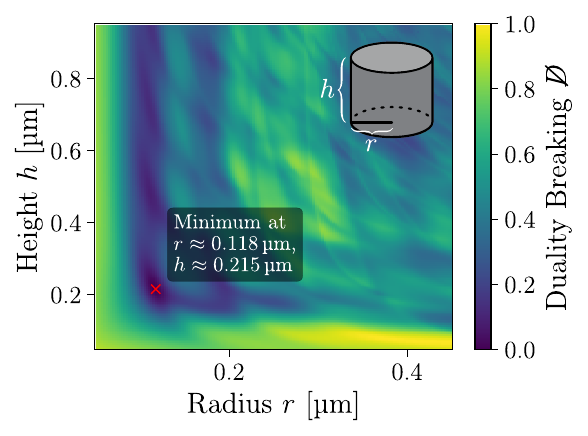}
\caption[]{Duality breaking $\cancel{D}$ according to Eq. (\ref{eq:db}) for isolated cylinders with variable dimensions. The dielectric cylinder with material properties similar to silicon ($\varepsilon_\mathrm{r, cyl, lossless} = 11.9$, $\mu_\mathrm{r, cyl} = 1$) at the considered infrared vacuum wavelength of $\lambda_0 = \qty{1}{\micro\meter}$ is embedded in a homogeneous dielectric akin to silicon dioxide ($n=\sqrt{\varepsilon_\mathrm{r, bg}}=1.44$, $\mu_\mathrm{r, bg} = 1$). The underlying T-matrices were retrieved from the T-matrix database accompanying Ref.~\citep{asadovaTmatrixRepresentationOptical2024}. Within the considered parameter ranges the cylinder geometry that minimizes $\cancel{D}$ is identified as $(r, h) \approx (\qty{118}{\nano\meter}, \qty{215}{\nano\meter})$, which is quite convenient for nanofabrication, as it conforms to top-down lithographic structuring of silicon on insulator (SOI).}
\label{fig-cyl}
\end{figure}

To identify a dual-symmetric scatterer, we express its associated transition matrix (T-matrix)~\citep{watermanMatrixFormulationElectromagnetic1965, beutelTreamsTmatrixbasedScattering2024}, which relates the incident and scattered field expanded in vector spherical harmonics, in basis states of well-defined helicity. To measure how much duality is broken by the scatterer, we quantify the duality breaking $\cancel{D}$ as~\citep{fernandez-corbatonDualChiralObjects2015}:

\begin{equation}
\label{eq:db}
\cancel{D} = \frac{\sum_\nu \sum_{\nu^\prime} \sum_{\lambda} | t_{\nu^\prime(-\lambda), \nu\lambda} |^2}{\sum_\nu \sum_{\nu^\prime} \sum_{\lambda} | t_{\nu^\prime(-\lambda), \nu\lambda} |^2 + | t_{\nu^\prime\lambda, \nu\lambda} |^2} \, .
\end{equation}
Here $t_{\nu^\prime\lambda^\prime, \nu\lambda}$ is the transition coefficient from $\ket{\nu\lambda}$ to $\ket{\nu^\prime\lambda^\prime}$, where we use the notation $\ket{\nu\lambda}$ for the basis state characterized by its helicity eigenvalue $\lambda$, while all remaining indices, such as energy, multipole degree and order are contracted into $\nu$. The duality breaking $\cancel{D}$ can be understood as the angle averaged probability of a photon's helicity flipping upon interacting with the scatterer. $\cancel{D}$ is limited to a range from zero to one by construction, with one indicating complete helicity flipping. A value of zero marks perfectly dual-symmetric scatterers, which exhibit equal scattering for corresponding electric and magnetic multipoles. Any radiatively coupled arrangement of such scatterers is also perfectly dual-symmetric~\citep{rahimzadeganCoreShellParticlesBuilding2018}. Even scatterers with finite $\cancel{D}\approx 0$ can then be used as building blocks to construct composite devices, that tend to maintain the approximate duality symmetry of their constituents~\citep{rahimzadeganCoreShellParticlesBuilding2018} (Provided that the remaining unmatched multipole coefficients of the constituent scatterers are not enhanced by the interaction, \textit{e.g.}, due to close proximity).

%  Designing scatterers to minimize their duality breaking generally benefits from unlocking additional degrees of freedom, as these enable matching the electric and magnetic transition coefficients for an increasing number of multipoles [@rahimzadeganCoreShellParticlesBuilding2018].

In Figure~\ref{fig-cyl}, a parameter sweep of the radius and height of a highly refractive dielectric cylinder, enclosed in some embedding material, is used to identify the most suitable configuration within the given design space. The Figure shows, that duality is optimally preserved for a radius $r\approx \qty{118}{nm}$ and height $h\approx \qty{215}{nm}$ with a remaining duality breaking of $\cancel{D}<3\times10^{ -3}$. The materials and operating wavelength have been chosen to resemble silicon and silicon dioxide in the near infrared~\citep{staudeMetamaterialinspiredSiliconNanophotonics2017}. For such cylinders, a large collection of precomputed T-matrices are available in the T-matrix database accompanying Ref.~\citep{asadovaTmatrixRepresentationOptical2024}.

%  While the materials used in the T-matrix database (and consequently in [Fig. %s](#fig-cyl)) were lossless, we have resimulated the T-matrices of the selected cylinder-geometry to add varying levels of dissipative loss ($\varepsilon_\mathrm{r, cyl} = 11.9 + \ii \varepsilon^{\prime \prime}$), to later illustrate the effects of nonidealities. To enable a computationally efficient, dense frequency sampling we leverage a sample-based pole-expansion of the T-matrix [@fischbachPoleExpansionTMatrixBased2026], which will allow us to clearly resolve the narrow qBIC line shapes below.

%  TODO switch to epsilon'' (or loss tangent)

\section{Metasurface with coincident BICs}

\begin{figure}[!htbp]
\centering
\includegraphics[width=1\linewidth]{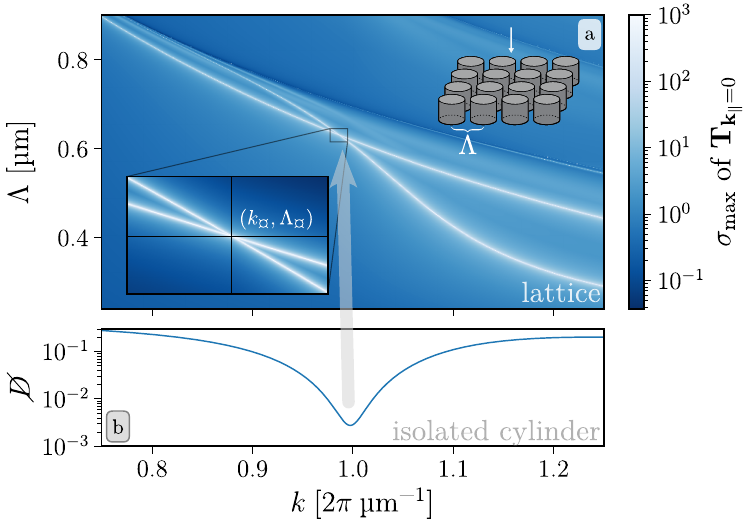}
\caption[]{Coincident BICs close to the point of minimum duality breaking. (a) Singularities of the lattice T-matrix $\mathrm{T}_{\mathbf{k}_\parallel=0}$ visualized by its largest singular value $\sigma_\mathrm{max}$. (b) Spectrally resolved duality breaking of the single cylinder.}
\label{fig-tune}
\end{figure}

Let us go ahead and place the chosen meta-atoms on a lattice, which will form two bound states in the continuum (BICs) that can be aligned spectrally by tuning the lattice constant. Once such coincident BICs are achieved, we will investigate the metasurface's response under plane wave illumination, highlighting the unusual helicity-preserving reflection unlocked by this configuration.

%  In contrast to [Fig. %s](#fig-cyl), the T-matrix of the isolated cylinder was resimulated using the commercially available finite element solver JCMwave to add material losses to the cylinders ($\varepsilon_\mathrm{r, cyl} = 11.9 + 10^{-4}\ii$). A pole-expansion of the T-matrix based on the AAA-algorithm for rational approximation was used to obtain a representation, that could be evaluated with arbitrary frequency resolution [@fischbachPoleExpansionTMatrixBased2026]. The lattice T-matrix was then computed using the open source solver `treams` [@beutelTreamsTmatrixbasedScattering2024].

To form a metasurface the selected cylinder is placed on a square lattice with periodicity $\Lambda$. The open source software package \texttt{treams} is used to study the optical response.
As a result of the lattice interactions, resonances with infinite radiative lifetime emerge at normal incidence, which are known as symmetry-protected bound states in the continuum (BICs). Their eigenenergy lies within the radiation continuum, from which they are decoupled, however, due to incompatible symmetries~\citep{hsuBoundStatesContinuum2016, sadrievaMultipolarOriginBound2019}. Here we are particularly interested in two such BICs, one with a mode-field dominated by the z-oriented electric dipole and one with a dominant z-oriented magnetic dipole contribution.
Because of the almost equal z-oriented electric and magnetic dipole polarizabilities at the design wavelength (guaranteed by the low duality breaking of the chosen scatterer), tuning one of the BICs to the design wavelength causes the other BIC to approach the design wavelength simultaneously~\citep{evlyukhinPolarizationSwitchingElectric2021, fischbachPoleExpansionTMatrixBased2026}. From Figure~\ref{fig-tune}(a), necessary lattice constant can be extracted at which the two BICs coincide. In particular, the colormap shows the largest singular value $\sigma_\mathrm{max}$ of the lattice T-matrix $\mathrm{T}_{\mathbf{k}_\parallel=0}$, which incorporates all multiple scattering effects between lattice sites at normal incidence~\citep{beutelEfficientSimulationBiperiodic2021} (\textit{i.e.}, for the Bloch wavevector $\mathbf{k}_\parallel=0$). The singularities of $\sigma_\mathrm{max}$ indicate resonances of the system~\citep{ustimenkoSingularValueDecomposition2026, fischbachPoleExpansionTMatrixBased2026}. In this case the BICs are visible as two white lines, which cross close to the design wavelength, where the duality breaking of the single cylinder is lowest (see Figure~\ref{fig-tune}(b)).

%  Here, we focus on two such BICs with TM and TE polarizations that emerge at the $\varepsilon^{\prime \prime}$-point of the considered metasurface. These states transform according to the irreducible representations $A_{\rm 2u}$ and $A_{\rm 2g}$ of the $D_{\rm 4h}$ symmetry group, respectively. The multipolar content of the TM-polarized BIC consists exclusively of vector spherical waves with the same symmetry, primarily the axisymmetric electric dipole (oriented along the z axis) and the axisymmetric electric octupole, while the TE-polarized BIC only contains the magnetic counterparts of these multipoles [https://doi.org/10.1103/PhysRevB.102.075103]. Importantly, the spectral locations of these BICs coincide when the electric and magnetic dipole polarizabilities of a single scatterer are equal [20, 23]. In our case, this condition is almost fulfilled due to the low duality breaking of the chosen scatterer. Moreover, while Refs. [20] and [23] consider this effect only within the dipolar approximation, here we shall also account for the contribution of higher-order multipoles, specifically the octupoles, by employing the T-matrix of a single scatterer truncated at least at $l_{\rm max} = 3$.

\begin{figure}[!htbp]
\centering
\vspace{-0.85cm}\includegraphics[width=0.85\linewidth]{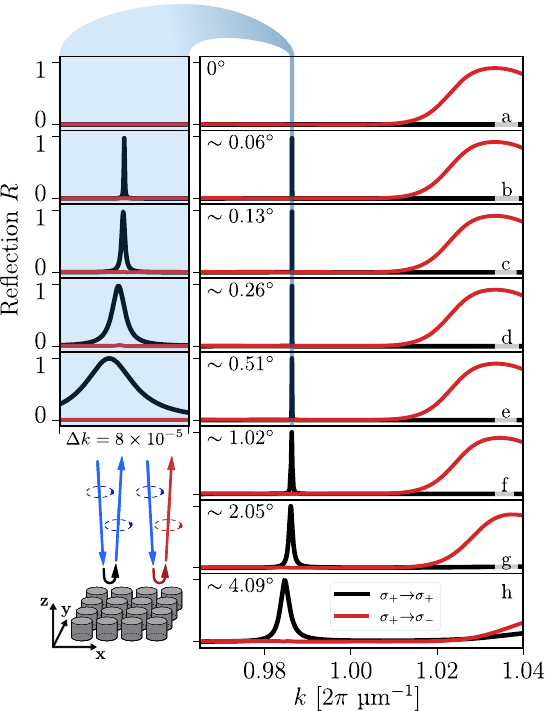}
\caption[]{Reflectance from cylinders on a square lattice under illumination by left circularly polarized plane waves at various angles of incidence, with suitably adjusted lattice constant $\Lambda$ (detailed in the appendix). The reflection under right circularly polarized illumination shows the same features and has been omitted for clarity. All panels show both the helicity preserving (black lines) and helicity flipping (red lines) reflection. These different modes of reflection are illustrated schematically in the lower left corner. The narrow column on the left shows zoomed-in frequency ranges of fixed width ($\Delta k = \qty{8e -5}{\per\micro\meter}$) around the resonances.}
\label{fig-refl}
\end{figure}

When illuminating this metasurface at normal incidence, the incident wave does not couple to the BICs. Figure~\ref{fig-refl}(a) shows the resulting reflection spectrum for different incident plane waves, which contains no visible resonance at the spectral location of the BICs. In addition, and as expected from the symmetry considerations above, the four-fold rotational symmetry combined with the low duality breaking of the single cylinders leads to a suppression of any backreflection around the design wavelength $\lambda_0 = 1 \unit{\per\micro\meter}$, which corresponds to an angular wavenumber of $k = 2\pi\unit{\per\micro\meter}$. Helicity-flipping reflection only starts to emerge at $k\gtrsim 1.02 \cdot 2\pi\unit{\per\micro\meter}$ where $\cancel{D}$ of the isolated cylinder increases, as the increasingly broken duality symmetry no longer guarantees helicity preservation. When a slight angle is introduced to the illumination, the rotational symmetry is broken while the duality symmetry is virtually unaffected. The two BICs transform into quasi-BICs (qBICs), as they start coupling to TM- and TE-polarized plane waves, respectively. The resonance frequency of the two qBICs has a distinct angle dependence. To keep the qBICs aligned with one another we have adjusted the lattice constant depending on the incidence angle. A detailed overview of the chosen lattice constants is provided in the appendix. In Figure~\ref{fig-refl}(b-h) we plot the helicity preserving (black lines) and helicity flipping reflectance (red lines) under illumination by a circularly polarized plane wave for a sequence of incidence angles. We clearly observe near-unity helicity-preserving reflection emerge at the frequency of the qBIC excitations, demonstrating a large change in the system's response even when the rotational symmetry is only slightly broken.

\begin{figure}[!htbp]
\centering
\includegraphics[width=1\linewidth]{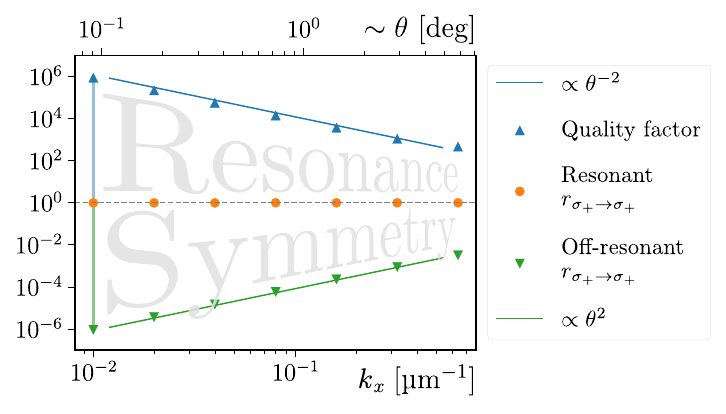}
\caption[]{Scaling of helicity-preserving reflection coefficients and qBIC quality factor (blue triangles) close to normal incidence without material losses. The reflection coefficient is evaluated at the resonance peak (orange dots) and at a fixed off-resonant wavelength of $\lambda_0=\qty{1}{\micro\meter}$ (green triangles). The blue and green solid lines illustrate the scaling behavior of the off-resonant reflection coefficient and the quality factor of the qBICs with asymmetry (in this case due to the angled incidence), which leads to the persistence of the perfect helicity-preserving reflection at small angles. The balance of symmetry suppression and resonant enhancement is schematically indicated by arrows.}
\label{fig-res-sym}
\end{figure}

Disregarding parasitic losses, the resonances persist at infinitesimal angles providing perfect helicity-preserving reflection (orange circles in Figure~\ref{fig-res-sym} remain at full reflection for small incidence angles), despite the discrete rotational symmetry being restored in this limit. This observation is in contrast to previous studies~\citep{abdelrahmanEffectsSymmetrybreakingElectromagnetic2021}, showing that without the qBICs the effect of breaking the discrete rotational symmetry is much less pronounced. To appreciate the typical suppression of helicity-preserving reflection close to normal incidence without the qBICs, we consider the reflection at an off-resonant wavelength. The magnitude of this off-resonant reflection coefficient is indicated by green triangles in Figure~\ref{fig-res-sym}. The reflectance is shown as a function of the lateral wave vector on a logarithmic scale, to better capture the response close to the symmetry condition. The reflectance is suppressed quadratically for small incidence angles. At the same time, the quality factor of the qBICs (blue triangles) follows the well known $1/\alpha^2$ scaling law~\citep{koshelevAsymmetricMetasurfacesHigh$Q$2018, kutuzovaQualityFactorScaling2023}, where the asymmetry parameter $\alpha$ in our case is the incidence angle $\theta$ (or equivalently the magnitude of the wavevector component parallel to the metasurface $k_x$). The $1/\theta^2$ scaling is observed because the qBICs are dominated by the z-oriented electric and magnetic dipoles, respectively, and their farfield decays linearly for small angles from the dipole axis. From temporal coupled mode theory it is known that the quality factor (\textit{i.e.}, the inverse radiative lifetime) scales quadratically with the farfield coupling coefficients~\citep{fanTemporalCoupledmodeTheory2003} leading to the observed $1/\theta^2$ scaling behavior. The previously discussed quadratic suppression of the off-resonant helicity-preserving reflection has a closely related origin: The helicity-preserving backscattering by a single off-axis cylinder is mediated predominantly by the tilted dipole polarizability. The power of two arises because the scattering process involves coupling to the polarizability from the farfield and subsequently radiating back to the farfield. When arranged in a weakly interacting lattice, this quadratic suppression of helicity-preserving backscattering persists. For the particular arrangement we present here, however, the configurational resonances (\textit{i.e.}, qBICs) eventually compensate the suppression, resulting in complete helicity-preserving reflection even for an infinitesimal symmetry breaking.

%  The scaling corresponds to the scaling of the squared far field amplitude of the z-oriented dipole field (*i.e.*, the vector spherical harmonic $\mathbf{X}(\theta, \varphi)$) and its inverse which are shown as green and blue solid lines, respectively.

\section{Conclusion}

In this article, we address symmetries that shape electromagnetic scattering. Understanding the consequences of intact and perturbatively broken symmetries has proven a valuable tool to guide the design of metasurfaces and other photonic devices~\citep{joannopoulosPhotonicCrystalsMolding2011, abdelrahmanEffectsSymmetrybreakingElectromagnetic2021, evlyukhinPolarizationSwitchingElectric2021, fernandez-corbatonDualChiralObjects2015, rahimzadeganCoreShellParticlesBuilding2018}. To adopt the learnings from symmetries to an imperfect reality, it is often assumed that perturbatively broken symmetries still strongly suppress phenomena that are forbidden in the ideal case. In contrast to this assumption, we demonstrate that four-fold rotationally symmetric metasurfaces composed of almost dual-symmetric meta-atoms exhibit a rapid emergence of strong helicity-preserving reflection at near-normal incidence, when they are designed to host an electric and a magnetic quasi-bound state in the continuum at the same frequency. Besides these findings being of fundamental scientific interest, they also have direct consequences for the development of tangible technology like helicity preserving cavities for enantio-selective detection.

The datasets generated and/or analyzed during the current study are available in the github repository, \href{http://www.github.com/tfp-photonics/resym}{www.github.com/tfp-photonics/resym}. Further the T-matrices generated during the current study are available in the T-matrix database accompanying reference~\citep{asadovaTmatrixRepresentationOptical2024}.

\vspace{1cm}
\section{Appendix}

\begin{figure}[!htbp]
\centering
\includegraphics[width=1\linewidth]{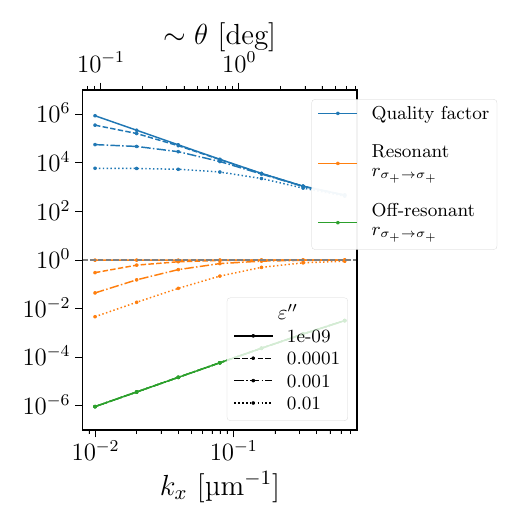}
\caption[]{Scaling of helicity preserving reflection and Q-factors in the presence of material losses. The shown quantities are the same as in Figure~\ref{fig-res-sym}. Different linestyles indicate varying levels of material loss $\varepsilon ^ {\prime \prime}$, which dampens the qBIC resonances leading to a limiting non-radiative Q-factor. As a result, the helicity preserving reflection starts decaying once the Q-factor limit is approached. Nonetheless, even for realistic levels of loss, the helicity preserving reflection remains largely intact up incidence angles of a few tenths of a degree.}
\label{fig-loss}
\end{figure}

As stated in the main text, the resonance frequencies of the two qBICs (one dominated by the z-oriented electric dipole, the other dominated by the z-oriented magnetic dipole), have different sensitivities to the incidence angle. To make sure both resonances overlap nonetheless, which will eventually generate the sought-after helicity-preserving reflection, the lattice constant was adjusted accordingly. The corresponding pairs of angles (alternatively expressed in terms of the x-component of the wavevector $k_x$) and lattice constants $\Lambda$ are provided in Table~\ref{tab-adjustlattice}.

\begin{table}
\centering
\caption[]{Adjusted lattice constants for the different samples considered in Figure~\ref{fig-refl} and Figure~\ref{fig-res-sym}. $\Lambda$ is provided to sufficient accuracy to ensure a good overlap between both qBICs.}
\label{tab-adjustlattice}
\begin{tabular}{p{\dimexpr 0.333\linewidth-2\tabcolsep\relax}p{\dimexpr 0.333\linewidth-2\tabcolsep\relax}p{\dimexpr 0.333\linewidth-2\tabcolsep\relax}}
\toprule
Angle [deg] & $k_x$ [$\unit{\per\micro\meter}$] & $\Lambda$ [$\lambda_0$] \\
\hline
0 & 0 & 0.62973864 \\
$\sim 0.06$ & 0.01 & 0.6297335 \\
$\sim 0.013$ & 0.02 & 0.629718 \\
$\sim 0.26$ & 0.04 & 0.62966 \\
$\sim 0.51$ & 0.08 & 0.6294 \\
$\sim 1.02$ & 0.16 & 0.6284 \\
$\sim 2.05$ & 0.32 & 0.6246 \\
$\sim 4.09$ & 0.64 & 0.6125 \\
\bottomrule
\end{tabular}
\end{table}

\subsection{Temporal coupled mode theory: qBICs without parasitic losses reach full reflection}

Let us start from Reference~\citep{fanTemporalCoupledmodeTheory2003} which assumes radiative loss to be the only relevant loss channel. Eq. (17) of Reference~\citep{fanTemporalCoupledmodeTheory2003} is valid for z-symmetric structures which support even (-) and odd (+) modes:
\begin{equation}
R = \frac{r^2 \delta^2 + t^2(1/\tau)^2 \mp 2rt\delta(1/\tau)}{\delta^2 + (1/\tau)^2}
\end{equation}

Here, we have introduced $\delta = \omega - \omega_0$ for a more compact notation. $r$ and $t$ are the reflection and transmission of the ``direct transport'' process, \textit{i.e.}, of the non-resonant background system. For a system without intrinsic losses, these obey $r^2 + t^2 = 1$ (stated just below Eq. (15) of Reference~\citep{fanTemporalCoupledmodeTheory2003}). Now let us show that there exists a $\delta$ such that $R=1$

\begin{equation}
\begin{aligned}
\Leftrightarrow && \underbrace{(r^2-1)}_{-t^2} \delta^2 + \underbrace{(t^2-1)}_{-r^2}/\tau^2 \mp 2rt\delta/\tau &= 0\\
\Leftrightarrow && (t\delta \pm r/\tau)^2 &= 0\\
\Leftrightarrow && \delta &= \mp \frac{r}{t\tau}
\end{aligned}
\end{equation}

Therefore, iff $t \neq 0$, a detuning $\delta$ exists for which the metasurface is fully reflective. For the metasurface treated in the main text, the generalized Kerker condition ensures suppressed reflection in the non-resonant case (\textit{i.e.}, $r\approx0$), which leads to the extremal reflection occurring close to the resonance frequency (i.e at $\delta\approx0$).

\subsection{Effect of material losses}

Depending on the intrinsic losses, such as material losses, losses due to surface imperfections, and leakage due to the finite size of the metasurface, the temporal coupled mode theory condition for reaching full reflection from the qBICs is no longer met. Considering an effective material loss $\varepsilon^{\prime \prime}$ in the cylinders, we find that the resonant reflection starts decaying for small angles of incidence (see Figure~\ref{fig-loss}). This behavior arises, because the light stored in the qBICs is absorbed rather than coupled out again to form the desired reflection. As a consequence, the observed helicity preserving reflection at small incidence angles only constitutes a strictly discontinuous limit in the absence of losses. Nonetheless, even in the presence of realistic material losses, the incidence angle up to which large helicity preserving reflection is observed is smaller than a degree.
\begin{acknowledgments}
J.D.F. and L.R. acknowledge support from the Karlsruhe School of Optics and Photonics (KSOP).
J.D.F. and C.R. acknowledge financial support by the Helmholtz Association in the framework of the innovation platform ``Solar TAP''.
L.R., I.F.C. and C.R. acknowledge financial support by the Deutsche Forschungsgemeinschaft (DFG, German Research Foundation) -- Project-ID 258734477 -- SFB 1173.
M.N. and C.R. acknowledge support by the KIT through the Virtual Materials Design (Virtmat) project.
N.U. and C.R. acknowledge financial support by the Deutsche Forschungsgemeinschaft (DFG, German Research Foundation) under Germany's Excellence Strategy via the Excellence Cluster 3D Matter Made to Order (EXC-2082/2, Grant No. 390761711) and from the Carl Zeiss Foundation via CZF-Focus@HEiKA.
\end{acknowledgments}
\bibliography{main.bib}

@book{weinbergQuantumTheoryFields1995,
	author = {Weinberg, Steven},
	year = {1995},
	publisher = {Cambridge university press},
	title = {The {Quantum} {Theory} of {Fields}},
	url = {https://books.google.com/books?hl=en%5C&lr=%5C&id=48xXMF1oHxkC%5C&oi=fnd%5C&pg=PR17%5C&dq=The+quantum+theory+of+fields%5C&ots=nlhvZkZ4%5C_F%5C&sig=ElpmwQ%5C_fxwa7rUzkgaaG6BNqe1M},
	volume = {2},
}

@inbook{noetherInvarianteVariationsprobleme1983,
	address = {Berlin, Heidelberg},
	author = {Noether, Emmy},
	booktitle = {Gesammelte {Abhandlungen} - {Collected} {Papers}},
	doi = {10.1007/978-3-642-39990-9_13},
	isbn = {978-3-642-39683-0 978-3-642-39990-9},
	year = {1983},
	pages = {231--239},
	publisher = {Springer Berlin Heidelberg},
	title = {Invariante {Variationsprobleme}},
}

@book{wignerGroupTheoryIts2012,
	author = {Wigner, Eugene},
	year = {1959},
	publisher = {Academic press},
	title = {Group {Theory}: And {Its} {Application} to the {Quantum} {Mechanics} of {Atomic} {Spectra}},
	url = {https://books.google.com/books?hl=en%5C&lr=%5C&id=ENZzI49uZMcC%5C&oi=fnd%5C&pg=PP1%5C&dq=Group+theory+and+its+application+to+the+quantum+mechanics+of+atomic+spectra%5C&ots=0MZhuKZJZI%5C&sig=5ekBEgsZnCOsOmEQRammC0JjzDA},
	volume = {5},
}

@book{degennesPhysicsLiquidCrystals1993,
	author = {De Gennes, Pierre-Gilles and Prost, Jacques},
	number = {83},
	year = {1993},
	publisher = {Oxford university press},
	title = {The {Physics} of {Liquid} {Crystals}},
	url = {https://books.google.de/books?hl=en%5C&lr=%5C&id=0Nw-dzWz5agC%5C&oi=fnd%5C&pg=PA1%5C&dq=de+Gennes+%5C%2526+Prost+1993%5C&ots=k6qpkqirut%5C&sig=BNoS4WKmUL8MfU2KWyK2HA2kwBI},
}

@article{fernandez-corbatonDualChiralObjects2015,
	author = {Fernandez-Corbaton, Ivan and Fruhnert, Martin and Rockstuhl, Carsten},
	journal = {ACS Photonics},
	doi = {10.1021/ph500419a},
	number = {3},
	year = {2015},
	month = {3},
	pages = {376--384},
	publisher = {American Chemical Society},
	title = {Dual and {Chiral} {Objects} for {Optical} {Activity} in {General} {Scattering} {Directions}},
	volume = {2},
}

@article{kiselevTestingAccuracyConvergence2025,
	author = {Kiselev, Andrei and Martin, Olivier J. F.},
	journal = {Optics Continuum},
	doi = {10.1364/OPTCON.559261},
	issn = {2770-0208},
	number = {3},
	year = {2025},
	month = {3},
	pages = {633},
	title = {Testing the {Accuracy} and {Convergence} of {Scattering} {Calculations} {Using} {Lorentz} {Reciprocity}},
	volume = {4},
}

@article{jarlskogCommutatorQuarkMass1985,
	author = {Jarlskog, C.},
	journal = {Physical Review Letters},
	doi = {10.1103/PhysRevLett.55.1039},
	issn = {0031-9007},
	number = {10},
	year = {1985},
	month = {9},
	pages = {1039--1042},
	title = {Commutator of the {Quark} {Mass} {Matrices} in the {Standard} {Electroweak} {Model} and a {Measure} of {Maximal} {CP} {Nonconservation}},
	volume = {55},
}

@article{wolfensteinParametrizationKobayashiMaskawaMatrix1983,
	author = {Wolfenstein, Lincoln},
	journal = {Physical Review Letters},
	doi = {10.1103/PhysRevLett.51.1945},
	issn = {0031-9007},
	number = {21},
	year = {1983},
	month = {11},
	pages = {1945--1947},
	title = {Parametrization of the {Kobayashi}-{Maskawa} {Matrix}},
	volume = {51},
}

@article{abdelrahmanEffectsSymmetrybreakingElectromagnetic2021,
	author = {Abdelrahman, Mohamed Ismail and Slivina, Evgeniia and Rockstuhl, Carsten and Fernandez-Corbaton, Ivan},
	journal = {Scientific Reports},
	doi = {10.1038/s41598-020-80347-5},
	issn = {2045-2322},
	number = {1},
	year = {2021},
	month = {1},
	pages = {1721},
	publisher = {Nature Publishing Group},
	title = {Effects of {Symmetry}-{Breaking} on {Electromagnetic} {Backscattering}},
	volume = {11},
}

@article{calkinInvariancePropertyFree1965,
	author = {Calkin, M. G.},
	journal = {American Journal of Physics},
	number = {11},
	year = {1965},
	pages = {958--960},
	title = {An {Invariance} {Property} of the {Free} {Electromagnetic} {Field}},
	url = {https://ui.adsabs.harvard.edu/abs/1965AmJPh..33..958C/abstract},
	howpublished = {https://ui.adsabs.harvard.edu/abs/1965AmJPh..33..958C/abstract},
	volume = {33},
}

@book{tungGroupTheoryPhysics1985,
	author = {Tung, Wu-Ki},
	isbn = {978-9971-966-57-7},
	year = {1985},
	publisher = {World Scientific},
	title = {Group {Theory} in {Physics}},
}

@article{zwanzigerQuantumFieldTheory1968,
	author = {Zwanziger, Daniel},
	journal = {Physical Review},
	doi = {10.1103/PhysRev.176.1489},
	number = {5},
	year = {1968},
	month = {12},
	pages = {1489--1495},
	publisher = {American Physical Society},
	title = {Quantum {Field} {Theory} of {Particles} with {Both} {Electric} and {Magnetic} {Charges}},
	volume = {176},
}

@misc{fernandez-corbatonHelicityDualitySymmetry2015,
	author = {Fernandez-Corbaton, Ivan},
	doi = {10.48550/arXiv.1407.4432},
	year = {2015},
	month = {12},
	publisher = {arXiv},
	title = {Helicity and {Duality} {Symmetry} in {Light} {Matter} {Interactions}: Theory and {Applications}},
}

@article{fernandez-corbatonForwardBackwardHelicity2013,
	author = {Fernandez-Corbaton, Ivan},
	journal = {Optics Express},
	doi = {10.1364/OE.21.029885},
	issn = {1094-4087},
	number = {24},
	year = {2013},
	month = {12},
	pages = {29885--29893},
	publisher = {Optica Publishing Group},
	title = {Forward and {Backward} {Helicity} {Scattering} {Coefficients} for {Systems} with {Discrete} {Rotational} {Symmetry}},
	volume = {21},
}

@article{nymanDigitalTwinChiral2024a,
	author = {Nyman, Markus and Garcia-Santiago, Xavier and Krsti{\' c}, Marjan and Materne, Lukas and Fernandez-Corbaton, Ivan and Holzer, Christof and Scott, Philip and Wegener, Martin and Klopper, Wim and Rockstuhl, Carsten},
	journal = {Laser \& Photonics Reviews},
	doi = {10.1002/lpor.202300967},
	issn = {1863-8880, 1863-8899},
	number = {6},
	year = {2024},
	month = {6},
	pages = {2300967},
	title = {A {Digital} {Twin} for a {Chiral} {Sensing} {Platform}},
	volume = {18},
}

@article{feisHelicityPreservingOpticalCavity2020,
	author = {Feis, Joshua and Beutel, Dominik and K{\" o}pfler, Julian and Garcia-Santiago, Xavier and Rockstuhl, Carsten and Wegener, Martin and Fernandez-Corbaton, Ivan},
	journal = {Physical Review Letters},
	doi = {10.1103/PhysRevLett.124.033201},
	number = {3},
	year = {2020},
	month = {1},
	pages = {033201},
	publisher = {American Physical Society},
	title = {Helicity-{Preserving} {Optical} {Cavity} {Modes} for {Enhanced} {Sensing} of {Chiral} {Molecules}},
	volume = {124},
}

@misc{pozharkovaLowsymmetryLatticesNonchiral2026,
	author = {Pozharkova, Anastasia and Blokhin, Oleg and Dyakov, Sergey A. and Baranov, Denis G.},
	year = {2026},
	month = {6},
	title = {Low-{Symmetry} {Lattices} of {Non}-{Chiral} {Meta}-{Atoms} for {Resonant} {Handedness}-{Preserving} {Reflection}},
	url = {https://arxiv.org/abs/2606.12067v1},
	howpublished = {https://arxiv.org/abs/2606.12067v1},
}

@misc{asadovaTmatrixRepresentationOptical2024,
	author = {Asadova, Nigar and Achouri, Karim and Arjas, Kristian and Augui{\' e}, Baptiste and Aydin, Roland and Baron, Alexandre and Beutel, Dominik and Bodermann, Bernd and Boussaoud, Kaoutar and Burger, Sven and Choi, Minseok and Czajkowski, Krzysztof M. and Evlyukhin, Andrey B. and Fazel-Najafabadi, Atefeh and Fernandez-Corbaton, Ivan and Garg, Puneet and Globosits, David and Hohenester, Ulrich and Kim, Hongyoon and Kim, Seokwoo and Lalanne, Philippe and Ru, Eric C. Le and Meyer, J{\" o}rg and Mun, Jungho and Pattelli, Lorenzo and Pflug, Lukas and Rockstuhl, Carsten and Rho, Junsuk and Rotter, Stefan and Stout, Brian and T{\" o}rm{\" a}, P{\" a}ivi and Trigo, Jorge Olmos and Tristram, Frank and Tsitsas, Nikolaos L. and Vall{\' e}e, Renaud and Vynck, Kevin and Weiss, Thomas and Wiecha, Peter and Wriedt, Thomas and Yannopapas, Vassilios and Yurkin, Maxim A. and Zouros, Grigorios P.},
	doi = {10.48550/arXiv.2408.10727},
	year = {2024},
	month = {8},
	publisher = {arXiv},
	title = {T-{Matrix} {Representation} of {Optical} {Scattering} {Response}: Suggestion for a {Data} {Format}},
}

@article{watermanMatrixFormulationElectromagnetic1965,
	author = {Waterman, P.C.},
	journal = {Proceedings of the IEEE},
	doi = {10.1109/PROC.1965.4058},
	issn = {1558-2256},
	number = {8},
	year = {1965},
	month = {8},
	pages = {805--812},
	title = {Matrix {Formulation} of {Electromagnetic} {Scattering}},
	volume = {53},
}

@article{beutelTreamsTmatrixbasedScattering2024,
	author = {Beutel, Dominik and Fernandez-Corbaton, Ivan and Rockstuhl, Carsten},
	journal = {Computer Physics Communications},
	doi = {10.1016/j.cpc.2023.109076},
	issn = {0010-4655},
	year = {2024},
	month = {4},
	pages = {109076},
	title = {\textit{{Treams}} -- a {T}-matrix-based {Scattering} {Code} for {Nanophotonics}},
	volume = {297},
}

@article{rahimzadeganCoreShellParticlesBuilding2018,
	author = {Rahimzadegan, Aso and Rockstuhl, Carsten and Fernandez-Corbaton, Ivan},
	journal = {Physical Review Applied},
	doi = {10.1103/PhysRevApplied.9.054051},
	number = {5},
	year = {2018},
	month = {5},
	pages = {054051},
	publisher = {American Physical Society},
	title = {Core-{Shell} {Particles} as {Building} {Blocks} for {Systems} with {High} {Duality} {Symmetry}},
	volume = {9},
}

@article{staudeMetamaterialinspiredSiliconNanophotonics2017,
	author = {Staude, Isabelle and Schilling, J{\" o}rg},
	journal = {Nature Photonics},
	doi = {10.1038/nphoton.2017.39},
	issn = {1749-4893},
	number = {5},
	year = {2017},
	month = {5},
	pages = {274--284},
	publisher = {Nature Publishing Group},
	title = {Metamaterial-{Inspired} {Silicon} {Nanophotonics}},
	volume = {11},
}

@article{hsuBoundStatesContinuum2016,
	author = {Hsu, Chia Wei and Zhen, Bo and Stone, A. Douglas and Joannopoulos, John D. and Solja{\v c}i{\' c}, Marin},
	journal = {Nature Reviews Materials},
	doi = {10.1038/natrevmats.2016.48},
	issn = {2058-8437},
	number = {9},
	year = {2016},
	month = {7},
	pages = {16048},
	title = {Bound {States} in the {Continuum}},
	volume = {1},
}

@article{sadrievaMultipolarOriginBound2019,
	author = {Sadrieva, Zarina and Frizyuk, Kristina and Petrov, Mihail and Kivshar, Yuri and Bogdanov, Andrey},
	journal = {Physical Review B},
	doi = {10.1103/PhysRevB.100.115303},
	number = {11},
	year = {2019},
	month = {9},
	pages = {115303},
	publisher = {American Physical Society},
	title = {Multipolar {Origin} of {Bound} {States} in the {Continuum}},
	volume = {100},
}

@article{evlyukhinPolarizationSwitchingElectric2021,
	author = {Evlyukhin, Andrey B. and Poleva, Maria A. and Prokhorov, Alexei V. and Baryshnikova, Kseniia V. and Miroshnichenko, Andrey E. and Chichkov, Boris N.},
	journal = {Laser \& Photonics Reviews},
	doi = {10.1002/lpor.202100206},
	issn = {1863-8899},
	number = {12},
	year = {2021},
	pages = {2100206},
	title = {Polarization {Switching} {Between} {Electric} and {Magnetic} {Quasi}-{Trapped} {Modes} in {Bianisotropic} {All}-{Dielectric} {Metasurfaces}},
	volume = {15},
}

@misc{fischbachPoleExpansionTMatrixBased2026,
	author = {Fischbach, Jan David and Betz, Fridtjof and Rebholz, Lukas and Garg, Puneet and Frizyuk, Kristina and Binkowski, Felix and Burger, Sven and Hammerschmidt, Martin and Rockstuhl, Carsten},
	doi = {10.48550/arXiv.2602.18414},
	year = {2026},
	month = {2},
	publisher = {arXiv},
	title = {Pole-{Expansion} of the {T}-{Matrix} {Based} on a {Matrix}-{Valued} {AAA}-{Algorithm}},
}

@article{beutelEfficientSimulationBiperiodic2021,
	author = {Beutel, Dominik and Groner, Achim and Rockstuhl, Carsten and Fernandez-Corbaton, Ivan},
	journal = {JOSA B},
	doi = {10.1364/JOSAB.419645},
	issn = {1520-8540},
	number = {6},
	year = {2021},
	month = {6},
	pages = {1782--1791},
	publisher = {Optica Publishing Group},
	title = {Efficient {Simulation} of {Biperiodic}, {Layered} {Structures} {Based} on the {T}-matrix {Method}},
	volume = {38},
}

@misc{ustimenkoSingularValueDecomposition2026,
	author = {Ustimenko, Nikita and Fernandez-Corbaton, Ivan and Rockstuhl, Carsten},
	doi = {10.48550/arXiv.2602.15741},
	year = {2026},
	month = {2},
	publisher = {arXiv},
	title = {Singular {Value} {Decomposition} to {Describe} {Bound} {States} in the {Continuum} in {Periodic} {Metasurfaces}},
}

@article{koshelevAsymmetricMetasurfacesHigh$Q$2018,
	author = {Koshelev, Kirill and Lepeshov, Sergey and Liu, Mingkai and Bogdanov, Andrey and Kivshar, Yuri},
	journal = {Physical Review Letters},
	doi = {10.1103/PhysRevLett.121.193903},
	number = {19},
	year = {2018},
	month = {11},
	pages = {193903},
	publisher = {American Physical Society},
	title = {Asymmetric {Metasurfaces} with {High}-\textdollar{}{Q}\textdollar{} {Resonances} {Governed} by {Bound} {States} in the {Continuum}},
	volume = {121},
}

@article{kutuzovaQualityFactorScaling2023,
	author = {Kutuzova, Aleksandra A. and Rybin, Mikhail V.},
	journal = {Physical Review B},
	doi = {10.1103/PhysRevB.107.195108},
	issn = {2469-9950, 2469-9969},
	number = {19},
	year = {2023},
	month = {5},
	pages = {195108},
	title = {Quality {Factor} {Scaling} of {Resonances} {Related} to {Bound} {States} in the {Continuum}},
	volume = {107},
}

@article{fanTemporalCoupledmodeTheory2003,
	author = {Fan, Shanhui and Suh, Wonjoo and Joannopoulos, J. D.},
	journal = {Journal of the Optical Society of America A},
	doi = {10.1364/JOSAA.20.000569},
	issn = {1084-7529, 1520-8532},
	number = {3},
	year = {2003},
	month = {3},
	pages = {569},
	title = {Temporal {Coupled}-{Mode} {Theory} for the {Fano} {Resonance} in {Optical} {Resonators}},
	volume = {20},
}

@book{joannopoulosPhotonicCrystalsMolding2011,
	author = {Joannopoulos, John D. and Johnson, Steven G. and Winn, Joshua N. and Meade, Robert D.},
	isbn = {978-1-4008-2824-1},
	year = {2011},
	month = {10},
	publisher = {Princeton University Press},
	title = {Photonic {Crystals}: Molding the {Flow} of {Light} - {Second} {Edition}},
}

@article{yangElectromagneticDualityProtected2020,
	author = {Yang, Qingdong and Chen, Weijin and Chen, Yuntian and Liu, Wei},
	journal = {ACS Photonics},
	doi = {10.1021/acsphotonics.0c00555},
	number = {7},
	year = {2020},
	month = {7},
	pages = {1830--1838},
	publisher = {American Chemical Society},
	title = {Electromagnetic {Duality} {Protected} {Scattering} {Properties} of {Nonmagnetic} {Particles}},
	volume = {7},
}
\end{document}